\documentclass[useAMS,usenatbib,twcolumn,usegraphicx]{mn2e}

\newcommand{\gsim}{\, \raisebox{-0.8ex}{$\stackrel{\textstyle >}{\sim}$ }}
\newcommand{\lsim}{\, \, \raisebox{-0.8ex}{$\stackrel{\textstyle <}{\sim}$ }}

\newcommand{\beq}{\begin{equation}}
\newcommand{\eeq}{\end{equation}}
\newcommand{\beqar}{\begin{eqnarray}}
\newcommand{\eeqar}{\end{eqnarray}}

\title[The 3rd family of compact objects]  
{Quark matter with strong magnetic field and possibility of the third family of compact stars}
\author[H. Sotani \& T. Tatsumi]
{Hajime Sotani$^1$ \thanks{E-mail:sotani@yukawa.kyoto-u.ac.jp}
and
Toshitaka Tatsumi$^2$
\\
$^1$Division of Theoretical Astronomy, National Astronomical Observatory of Japan, 2-21-1 Osawa, Mitaka, Tokyo 181-8588, Japan\\
$^2$Department of Physics, Kyoto University, Kyoto 606-8502, Japan}

\begin{document}
\maketitle
\label{firstpage}

\begin{abstract}
We consider the possibility for the existence of the third family of compact objects, considering the effect of strong magnetic fields inside the hybrid stars. As a result, we demonstrate such new sequences of stable equilibrium configurations for some hadronic equations of state. Through the analysis of the adiabatic index inside stars, we find the conditions for appearing the third family of compact objects, i.e., for hadronic stars without quarks, that the maximum mass should be small, the central density for the maximum mass should be also small, and the radius for the the maximum mass should be large. Even for soft hadronic equations of state, the  two solar-mass stars might survive as the third family of compact objects, once quark matter with strong magnetic field, such as $\sim {\cal O}(10^{19} {\rm G})$, is taken into account. It might give a hint to solve the so-called hyperon puzzle in nuclear physics. 
\end{abstract}


\section{Introduction}
\label{sec:I}

Nowadays, properties of matter under the extreme conditions such as high-density or high-temperature region have been one of the most interesting subjects. Recent progress in observation of neutron stars has provided us of interesting results about the maximum mass, the magnetic field, surface temperature and so on. Theoretically, however, there is still large ambiguity about the equation of state (EOS) at high-density region.  In particular, the density inside the neutron stars can be well over the nuclear saturation density, which leads to the difficulty to determine the structures of neutron stars. Thus, via the observations of neutron stars itself and the phenomena associated with the neutron stars, one could obtain the information about such a high density region.

The asteroseismology is one of the most powerful techniques to extract the interior information (e.g., \cite{AK1996,STM2001,SKH2004,SYMT2011,DGKK2013}). The gravitational waves must be suitable observables for adopting the asteroseismology, although the direct observations of the gravitational waves have not been successful yet. Even excluding the gravitational waves, fortunately, there are evidences of observations of neutron star oscillations. That is the quasi periodic oscillations discovered in the afterglow of the giant flares. To explain the quasi periodic oscillations theoretically, there are many attempts in terms of the crustal torsional oscillations (e.g., \cite{SW2009,GNHL2011,Sotani2011,Sotani2014}) and/or the magnetic oscillations (e.g., \cite{Sotani2007,Sotani2008,Sotani2008b,Sotani2009,CK2011,PA2012,GCSFM2012,GCFMS2013}). Through such attempts, it is shown that, identifying the observed quasi periodic oscillations with the crustal torsional oscillations, one can constrain the EOS in the crust region \citep{SNIO2012,SNIO2013a,SNIO2013b,Sotani2016,SIO2016,SIO2017}.

On the other hand, discovery of the $2M_\odot$ neutron stars also can put a strong constraint on the EOS for neutron star matter \citep{D2010,A2013}, where $M_\odot$ denotes the solar mass. That is, the existence of these massive compact objects rules out several soft EOSs, with which the expected maximum mass of compact object is less than the observed maximum mass. Meanwhile, in nuclear physics, it is becoming a standard conception that hyperons would appear at the high density region such as $\sim (2-3)\times n_0$, where $n_0$ denotes the nuclear saturation density. But, the introduction of hyperons makes the EOS soft enough to be ruled out by the existence of $2M_\odot$ neutron stars. This is a serious problem in nuclear physics, which is the so-called hyperon puzzle. So far, there are several discussions in the astrophysical scenario in the context of the hyperon puzzle (e.g., \citet{takatsuka04,takatsuka08,Peres2013,Dalena2014,Oertel2015,Oertel2016,Fortin2016}).

Compact objects can be left after the death of massive stars. Depending on the mass of progenitors, the resultant compact stars are different. In general, two families of compact stars are well-known. One is white dwarfs, which are supported by the electron degeneracy pressure. Second is neutron stars, which are supported by the neutron degeneracy pressure together with the strong nuclear repulsion forces in the sort range \citep{shapiro-teukolsky}. The possibility for the existence of the third family of compact stars before the gravitational collapse to black holes has been an interesting and important issue in nuclear physics and astrophysics \citep{Glen2000,Haen2007}. One of the necessary conditions is related to the gravitational stability of superdense stars;
\begin{equation}
  \Gamma>\frac{4}{3}\left(1+K\frac{R_s}{R}\right),
  \label{cond1}
\end{equation}
in terms of the adiabatic index $\Gamma$ within general relativity, where $R$ is the stellar radius, $R_s=2MG/c^2$ is the Schwarzschild radius, $K$ is of order unity, and $M$ is the stellar mass.  Since the third family of compact objects should be the new stable sequence with the central density $\rho_{\rm c}$ beyond the maximum central density of neutron stars $\rho_{\rm c,max}$, this condition should be fulfilled for superdense stars with the central density $\rho_{\rm c}>\rho_{\rm c,max}$. However, this condition may not be sufficient for the existence of the third family. As another condition, early qualitative discussion of Gerlach has suggested that the third family is possible if the adiabatic index (or speed of sound) abruptly increases beyond $\rho_{\rm c,max}$ \citep{G1968}.

If there is a third family, it should be hybrid stars. 
In the seminal paper, \cite{Baym1977} has discussed the possibility of hybrid stars as the third family of compact stars, where he concluded that hybrid stars with $\rho_{\rm c}>\rho_{\rm c,max}$ are most likely unstable, because $\Gamma\rightarrow 4/3$ or the sound velocity $c_s^2\rightarrow 1/3$ as a limiting behavior of quark matter at high-density due to the asymptotic freedom.

In our previous paper, we have discussed the effect of the strong magnetic field on EOS of quark matter, and shown that $\Gamma\rightarrow 2$ within the bag model EOS \citep{ST2015}. Thus, we have pointed out that the maximum mass may well clear the recent observation of $2M_\odot$\footnote{The stiffening effect of the magnetic field on the EOS has been also studied in the context of the possibility of the super-Chandrasekhar white dwarfs \citep{Kun2012,Das2012,Das2013}.}. In this paper, we consider the possibility of the third family of compact stars in the same context. The EOS of quark matter can then satisfy Eq.~(\ref{cond1}). Taking Shen's EOS \citep{Shen-EOS} and modified Shen's EOSs \citep{ShenL-EOS,Ishizuka-EOS} as typical ones for neutron star matter, we will discuss whether the third family could exist or not, and such conditions if the third family exists. We remark that the Shen EOS is one of the EOSs adopted in many numerical simulations so far.

The magnetic field inside neutron stars has been an unsolved problem. In addition to the standard neutron stars, the existence of strongly magnetized neutron stars, the so-called magnetars, has been suggested theoretically \citep{DT1992,TD1993,TD1996} and observationally \citep{K1998,H1999,M1999}. However, the configuration and strength of magnetic fields inside the star are still quite uncertain. Using the virial theorem, the maximum limit of the magnetic field strength can be estimated to be $\sim 10^{18}-10^{19}$ G for the canonical neutron star model \citep{LS1991,CPL2001}. Note that such estimation is simply derived from the Newtonian virial theorem based on the assumption that the field strength inside the star is constant. On the other hand, the relativistic version of virial theorem is more complicated \citep{GB1994} and the possible maximum strength in relativistic case could be enhanced. The variation of the magnetic field inside stars should also enhance it.


Furthermore, the origin of the magnetic fields inside neutron stars is still uncertain. The simplest scenario might be the explanation due to the fossil magnetic field inherited from progenitor stars. That is, the small magnetic field would be amplified during the gravitational collapse with the conservation of magnetic flux, which leads to the strong magnetic field in neutron stars \citep{C1992}. The canonical magnetic field of neutron stars might be explained with this simple scenario, but the case of magnetars is quite difficult \citep{tatsumi00}. As another generation mechanism, the magnetohydrodynamic dynamo has been also suggested, where the rapidly rotating protoneutron star with the spin period smaller than 3 ms may amplify a seed of the magnetic field up to $\sim 10^{15}$ G \citep{DT1992,TD1993}, although this scenario seems to be unacceptable from the observations of supernova remnants related to the magnetar candidates \citep{VK2006}. Moreover, as an origin of a strong magnetic field inside neutron stars, one of the authors (T.T.) has suggested the possibility of  ferromagnetism of quark liquid inside the stars \citep{tatsumi00}.

In this paper, we consider the strong magnetic field inside hybrid stars, which are composed primary of free quarks.
Up to now, it has been also discussed how the strong the magnetic field affects behavior of quark matter in high density region, e.g., strange quark star models in strong magnetic fields within a confining model \citep{Sinha2013}, self-consistent configuration of magnetic field in hybrid stars \citep{Franzon2016}, and effect of temperature on two-flavor superconducting quark matter with the magnetic field effect \citep{Mandal2016}. Meanwhile,  as an extreme case, we focus on the effect of the lowest Landau level on the EOS of quark matter in this paper.
In practice, it was shown that quarks can settle on only the lowest Landau level, if the magnetic field strength is much stronger than a critical field strength. As a result, the quark EOS becomes so stiff, and quark matter largely occupies inside the resultant hybrid star \citep{ST2015}. 
We especially examine the possibility for existence of the third family of compact objects by considering such hybrid stars. Unless otherwise noted, we adopt the geometric unit of $c=G=1$, where $c$ and $G$ denote the speed of light and the gravitational constant, respectively.

\section{Hybrid stars with strong magnetic field}
\label{sec:II}

We assume the hadron-quark deconfinement transition in the core region. The mechanism or properties of the deconfinement transition has large ambiguity at high-density region, while many studies have suggested it. To construct hybrid stars, the hadronic EOS have to be connected to the quark EOS in some wise. So far, in order to discuss the dependence of the stellar models on the EOSs, sometimes the EOS, i.e., $p(\varepsilon)$ with $p$ and $\varepsilon$ being pressure and the energy density, respectively,  are simply connected in the $p-\varepsilon$ plane (e.g., \cite{RR1974,STM2001,AMP2013,BS2015}). In the same way as in such previous works, in this paper, we simply connect the hadronic EOS with the quark EOS at the transition density $\varepsilon_c$ to avoid the complexity;
\begin{equation}
  p(\varepsilon)=\left\{
  \begin{array}{@{\,}cc@{\,}}
     p_H(\varepsilon) & \varepsilon < \varepsilon_c \\
     p_Q(\varepsilon) & \varepsilon \ge \varepsilon_c
  \end{array}\right. ,
  \label{hqeos}
\end{equation}
where $p_{H(Q)}$ denotes the pressure in the hadronic (quark) phase and $\varepsilon_c$ is defined by $p_H(\varepsilon_c)=p_Q(\varepsilon_c)$. We remark that the baryon number density becomes discontinuous at the interface between the hadronic and quark phases (see Table \ref{tab:nb}), i.e., this simple connection is thermodynamically inconsistent. If the deconfinement transition is of the first order, we must take into account the mixed phase by way of the Maxwell construction or more elaborated Gibbs construction \citep{Glen1992,Heis1993,Vosk2002,Maru2007}. Fortunately, we can check that the bulk properties of hybrid stars have little dependence on the details of the phase transition, at least in the maximum mass region \citep{ST2015}.

In this paper, we adopt four hadronic EOSs as shown in Table \ref{tab:NS-properties}, i.e., one EOS composed of only nucleons and three EOSs including the hyperons, in order to construct hybrid stars. Those are based on the relativistic mean field approach. \cite{Shen-EOS} have originally derived the EOS table composed of only nucleons in the wide ranges of density, temperature, and proton fraction, where they also used the Thomas-Fermi approximation to describe the heavy nuclei in the low density region. Afterward, they modified their EOS in the same framework as the original one, but the contribution of $\Lambda$ hyperons was also taken into account \citep{ShenL-EOS}. Hereafter, we refer to the original and modified EOSs as ``HShen EOS" and ``HShen $\Lambda$ EOS", respectively.

On the other hand, \cite{Ishizuka-EOS} have considered the contributions of $\Lambda$, $\Sigma$, and $\Xi$ hyperons with and without pions into the HShen EOS, where they adopted $U_{\Lambda}^{(N)}= -30$ MeV, $U_{\Sigma}^{(N)}=+30$ MeV, and $U_{\Xi}^{(N)}=-15$ MeV, as standard values of hyperon potentials at the saturation density. We remark that they consider the pion contributions to EOS, assuming that the pion interaction is neglected. Thus, the effect of introduction of pions may be oversimplified (see \cite{Ishizuka-EOS} for detail about the introduction of pion). Even so, as one of the hadronic EOSs, we also adopt such an EOS. Hereafter, we refer to the HShen EOS modified in \cite{Ishizuka-EOS} as ``HShen Y EOS" without the pion contribution and ``HShen Y$\pi$ EOS" with the pion contribution, respectively.

\begin{figure*}
\begin{center}
\begin{tabular}{cc}
\includegraphics[scale=0.5]{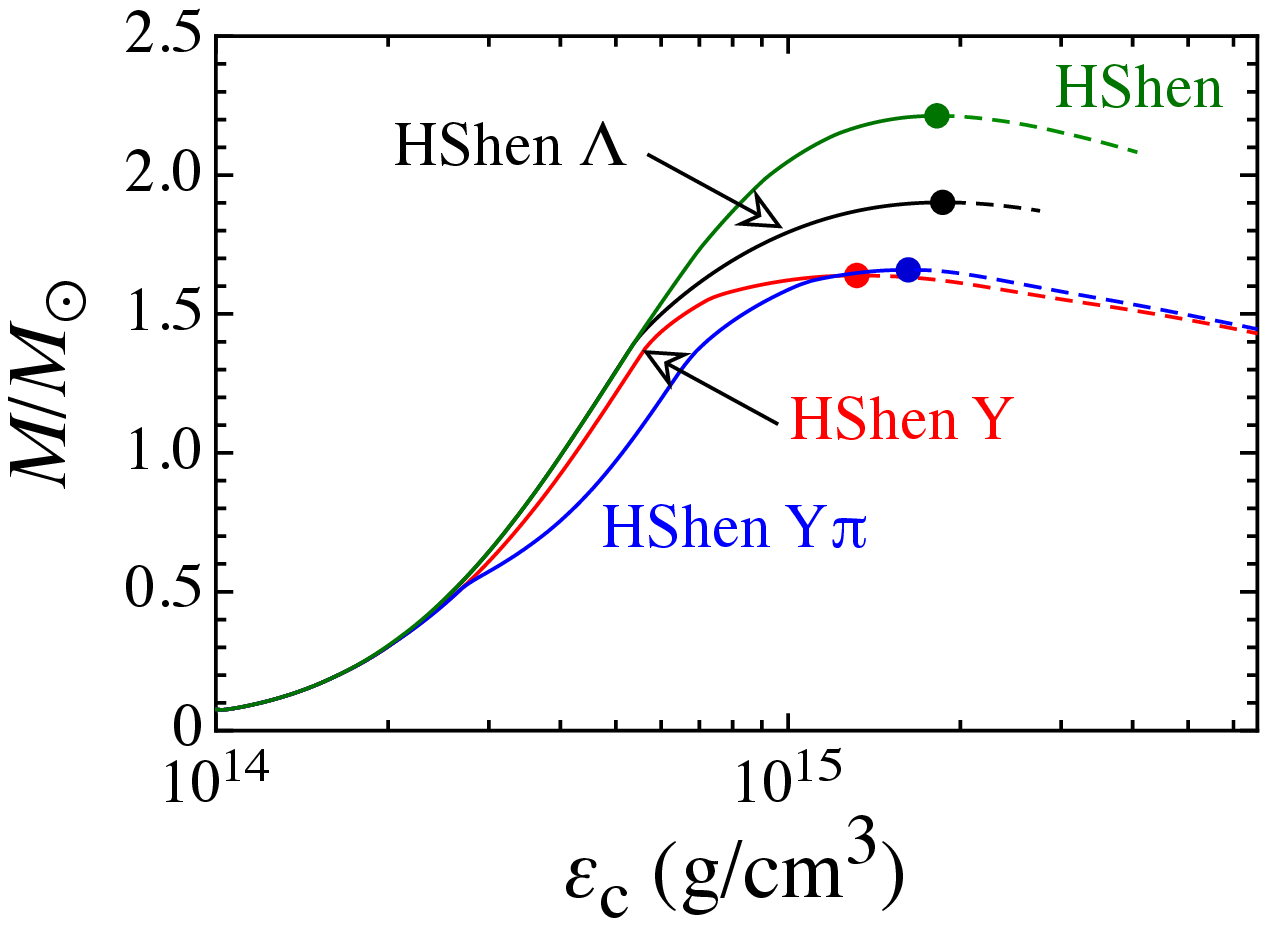} &
\includegraphics[scale=0.5]{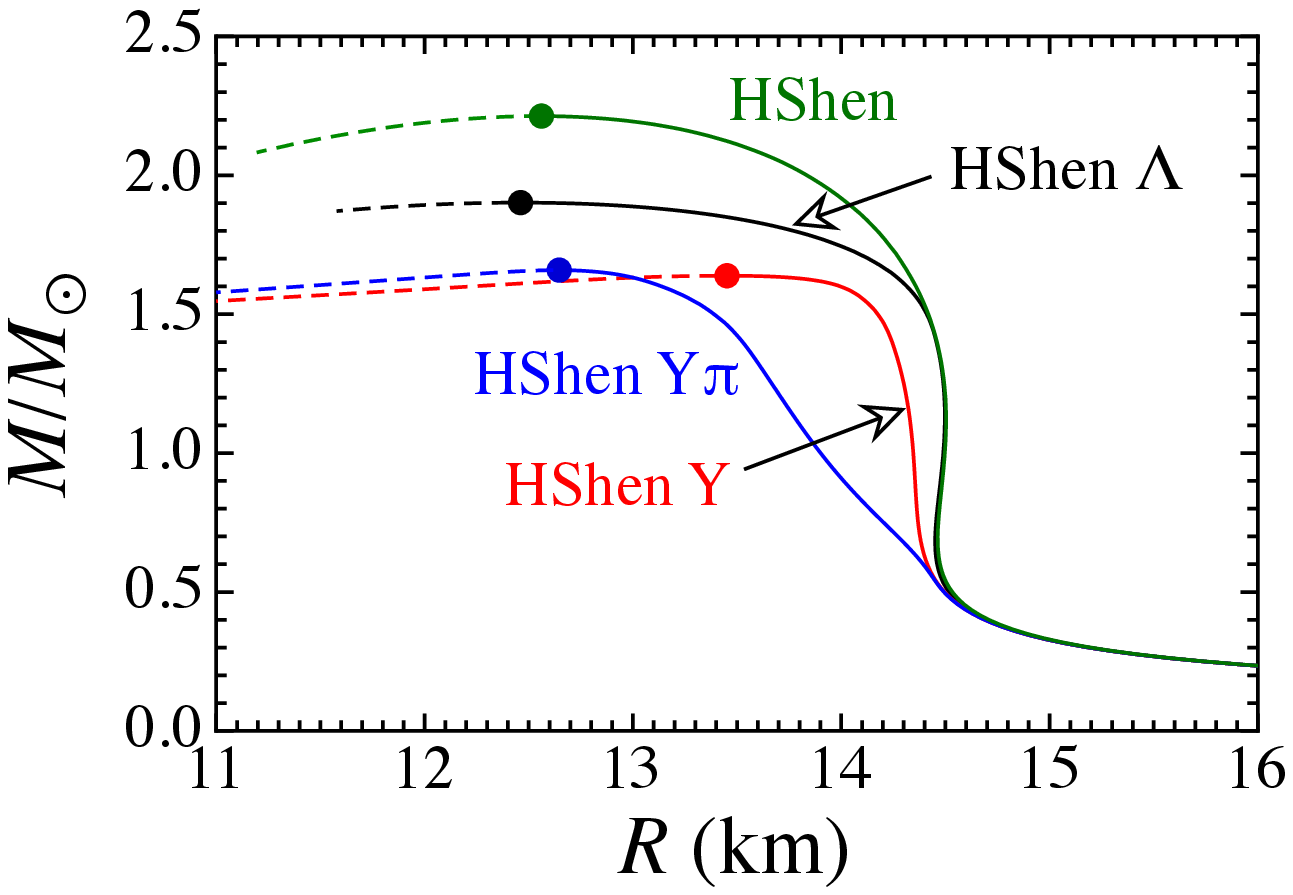}
\end{tabular}
\end{center}
\caption{
With the HShen, HShen $\Lambda$, HShen Y, and HShen Y$\pi$ EOSs, the neutron star mass is shown as a function of the central density in the left panel, while the mass-radius relations are shown in the right panel. In the both panels, the marks denote the maximum mass expected with each EOS. The solid and broken lines correspond to the stable and unstable stellar models, respectively. 
}
\label{fig:HShen}
\end{figure*}

Since three EOSs with hyperons considered here are improved HShen EOS by adding the contributions of hyperons, those become completely equivalent to the HShen EOS for the density region lower than the critical density where the hyperon and/or pion can appear. In Fig. \ref{fig:HShen}, we show the neutron star mass for each EOS as a function of the central density in the left panel and that as a function of stellar radius in the right panel. In the figure, the solid and broken lines correspond to the stable and unstable equilibrium models, respectively, and the marks denote the maximum mass expected for each EOS. The properties of neutron star model with maximum mass for each EOS are shown in Table \ref{tab:NS-properties}. From this figure, one can observe that the stellar masses constructed with EOSs with hyperons can not be over $2M_\odot$, which is the maximum mass observed so far, even though the original HShen EOS can construct the stellar model with the mass larger than $2M_\odot$. Hereafter, we mainly focus on the HShen and HShen Y EOSs to discuss the possibility of third family of compact objects. This is because, the HShen Y EOS is an example for the case where the third family can exist, while the HShen EOS is for the case where the third family cannot exist, as shown later.

\begin{table}
\centering
\caption{Expected maximum masses of neutron stars constructed with the hadronic EOSs without the effect of magnetic fields, and the corresponding stellar radii and the central densities. 
}
\begin{tabular}{lccc}
\hline\hline
EOS & $M_{\rm max}/M_\odot$ & $R$ (km) & $\varepsilon_{\rm c}$ (g/cm$^3$)  \\
\hline
HShen                    &   $2.21$ & $12.6$ & $1.82 \times 10^{15}$  \\
HShen $\Lambda$ &   $1.90$ & $12.5$ & $1.86 \times 10^{15}$  \\
HShen Y                &   $1.64$ & $13.5$ & $1.32 \times 10^{15}$ \\
HShen Y$\pi$        &   $1.66$ & $12.6$ & $1.62 \times 10^{15}$ \\
\hline\hline
\end{tabular}
\label{tab:NS-properties}
\end{table}

For the quark matter EOS we use here the bag model EOS for simplicity \citep{Far1984,ST2015}\footnote{It is a popular EOS for quark matter, minimally including the nonperturbative effect by the bag constant $\cal B$. EOS based on the Nambu-Jona-Lasinio model may be also available \citep{Men2009}, where the effective bag constant corresponding to $\cal B$ can be reduced from EOS. Basically, the both should share a similar feature at high density, where chiral symmetry is restored.}. The original form of the bag model EOS can be written as,
\begin{equation}
  p=\frac{1}{3}(\varepsilon-4{\cal B}),
\label{bagEOS}
\end{equation}
where ${\cal B}$ denotes the bag constant and the energy density $\varepsilon$ can be written with the baryon number density $n_{\rm b}$ as 
\begin{equation}
  \varepsilon=\frac{3}{4}\pi^{2/3}\hbar cn_{\rm b}^{4/3}+{\cal B}
\end{equation}
for SU(3) flavor symmetric massless quark matter.

As mentioned before, in order to construct the third family of compact objects, at least, one has to take into account the additional physics to support the gravity of compact objects. In this paper, as such an additional physics, we consider the strong magnetic fields inside stars \citep{ST2015}. With respect to the EOS, the effect of the magnetic fields should be more important in quark matter rather than in hadronic matter, because (i) the quark masses are much smaller than the baryon mass and (ii) all quarks have net electric charge while the neutrons are dominant in the hadronic phase. With such reasons, for simplicity, we consider the effect of the magnetic fields only on quark matter, as in \cite{ST2015}, where we omit the effect of magnetic field on the hadronic EOS.
In the following we assume the density dependent magnetic field inside stars, without recourse to its details \citep{Ban1997}, i.e., the weak magnetic field in the low density region and the strong magnetic field in the high density region.  An actual calculation has shown that the bulk properties of hadronic EOS is little affected up to $B={\cal O}(10^{17} {\rm G})$ \citep{Cas2014}. Thus, although for a realistic case one might see the effect of magnetic field even in the hadronic EOS with such a strong magnetic field as in this paper, as a first step we omit such effect and focus on the effect of magnetic field only in quark matter with the reasons mentioned the above.

In general, many Landau levels are occupied by quarks if the magnetic fields are not so strong. But, one may have to consider the effect of the Landau levels, if the strength of magnetic fields is strong enough for quarks to occupy only some lower levels. In particular, the extreme case is when quarks settle only in the lowest Landau level, which can realize if the strength of magnetic fields becomes $\sim {\cal O}(10^{19}$ G) \citep{ST2015}\footnote{Compered to the observations of field strength of magnetars such as $\sim 10^{15}$ G, the order of $\sim 10^{19}$ G may be much larger. However, it could be possible as in the case of white dwarf, i.e., the observed surface strength of magnetic field of white dwarf is at most in the range of $10^{6}-10^8$ G \citep{KC1990,K2013}, while the central field strength in the order of $10^{12}$ G can be theoretically constructed \citep{Ost1968}. As another possibility, such a strong magnetic field in core region may be screened out by the ambient high-conductivity plasma \citep{HS2012}.}. Assuming that the uniform magnetic field locally points toward a specific direction, that the quark mass does not contribute significantly to the energy level of quark matter, and that quark matter becomes flavor symmetric, the EOS for quark matter settling only in the lowest Landau level can be expressed within the MIT bag model, as
\begin{equation}
  p = \varepsilon - 2{\cal B},   \label{eq:eos}
\end{equation}
where the energy density $\varepsilon$ can be written as 
\begin{equation}
   \varepsilon=\frac{5\pi^2\hbar^2c^2}{2eB}n_b^2+{\cal B},
\label{qeose}
\end{equation}
\citep{ST2015}. Hereafter, we refer to this quark EOS as ``Landau EOS''. The relation between $p$ and $\varepsilon$ given by Eq. (\ref{eq:eos}) is satisfied independently of the magnetic field strength, when the strength would be larger than the critical strength, $B_c$. We remark that the Landau EOS is the limiting case of a stiff EOS, because the adiabatic sound speed, $c_{\rm s}=(dp/d\varepsilon)^{1/2}$,  becomes the speed of light $c$. We, hereafter, use the notation $\varepsilon_{\rm L}$ for the transition density $\varepsilon_{c}$ to specify the quark EOS. We also remark that the Landau EOS is independent, even if one simply considers the effects of magnetic field on EOS such as $p\to p+B^2/2$ and $\varepsilon\to \varepsilon + B^2/2$ (e,g,. \cite{Cas2014})\footnote{In more realistic case, one might see anisotropy in the magnetic pressure, depending on the magnetic field configuration. In addition, one might take into account the magnetization of the system (e.g., \cite{Das2012}).}.

In Table \ref{tab:nb}, for given the values of $\varepsilon_{\rm L}$, we show the corresponding values of the bag constant, critical magnetic field strength, baryon number density with the critical magnetic field in the quark phase, and baryon number density in the hadronic phase for the HShen and HShen Y EOSs. We also remark that there is still left a large uncertainty in the value of the bag constant. The MIT group has originally obtained ${\cal B}=57.5$ MeV/fm$^3$ \citep{MIT}. Subsequently, the various values of ${\cal B}$ are extracted by several groups via fitting to light hadron spectra, which reaches up to $\sim 351.7$ MeV/fm$^3$ \citep{CHP1983}. Recently, the possibility of ${\cal B}=O(500)$ MeV/fm$^3$ has been suggested by evaluating the energy-momentum tensor in QCD or using the data of the deconfinement temperature within the lattice QCD calculations \citep{B2005}. So, in this paper we consider in the wide range of ${\cal B}$, i.e., ${\cal B}\lsim 500$ MeV/fm$^3$, to discuss the existence of third family of compact objects.

\begin{table*}
\centering
\caption{
At the given the values of transition density $\varepsilon_{\rm L}$, the corresponding values of the bag constant ${\cal B}$, critical magnetic field strength $B_c$, baryon number density with the critical magnetic field in the quark phase $n_{\rm b}^{(Q)}$, and baryon number density in the hadron phase $n_{\rm b}^{(H)}$ for the HShen and HSnen Y EOSs.
}
\begin{tabular}{lccccc}
\hline\hline
EOS & $\varepsilon_{\rm L}$ (g/cm$^3$) & ${\cal B}$ (MeV/fm$^3$) & $B_c$ (G) & $n_{\rm b}^{(Q)}$ (fm$^{-3}$) & $n_{\rm b}^{(H)}$ (fm$^{-3}$) \\
\hline
HShen          &   $1.40\times 10^{15}$ & $284.5$ & $5.10 \times 10^{19}$ & $0.893$ & $0.671$ \\
                     &   $1.83\times 10^{15}$ & $352.5$ & $5.92 \times 10^{19}$ & $1.116$ & $0.825$ \\
                     &   $2.40\times 10^{15}$ & $442.2$ & $6.86 \times 10^{19}$ & $1.391$ & $1.012$ \\
\hline                     
HShen Y       &   $1.0\times 10^{15}$  &  $247.1$ & $4.04 \times 10^{19}$ & $0.629$ & $0.526$ \\
                     &   $1.4\times 10^{15}$  &  $339.9$ & $4.81 \times 10^{19}$ & $0.818$ & $0.710$ \\
                     &   $2.0\times 10^{15}$  &  $470.2$ & $5.82 \times 10^{19}$ & $1.088$ & $0.969$  \\
\hline\hline
\end{tabular}
\label{tab:nb}
\end{table*}

\begin{figure*}
\begin{center}
\begin{tabular}{cc}
\includegraphics[scale=0.5]{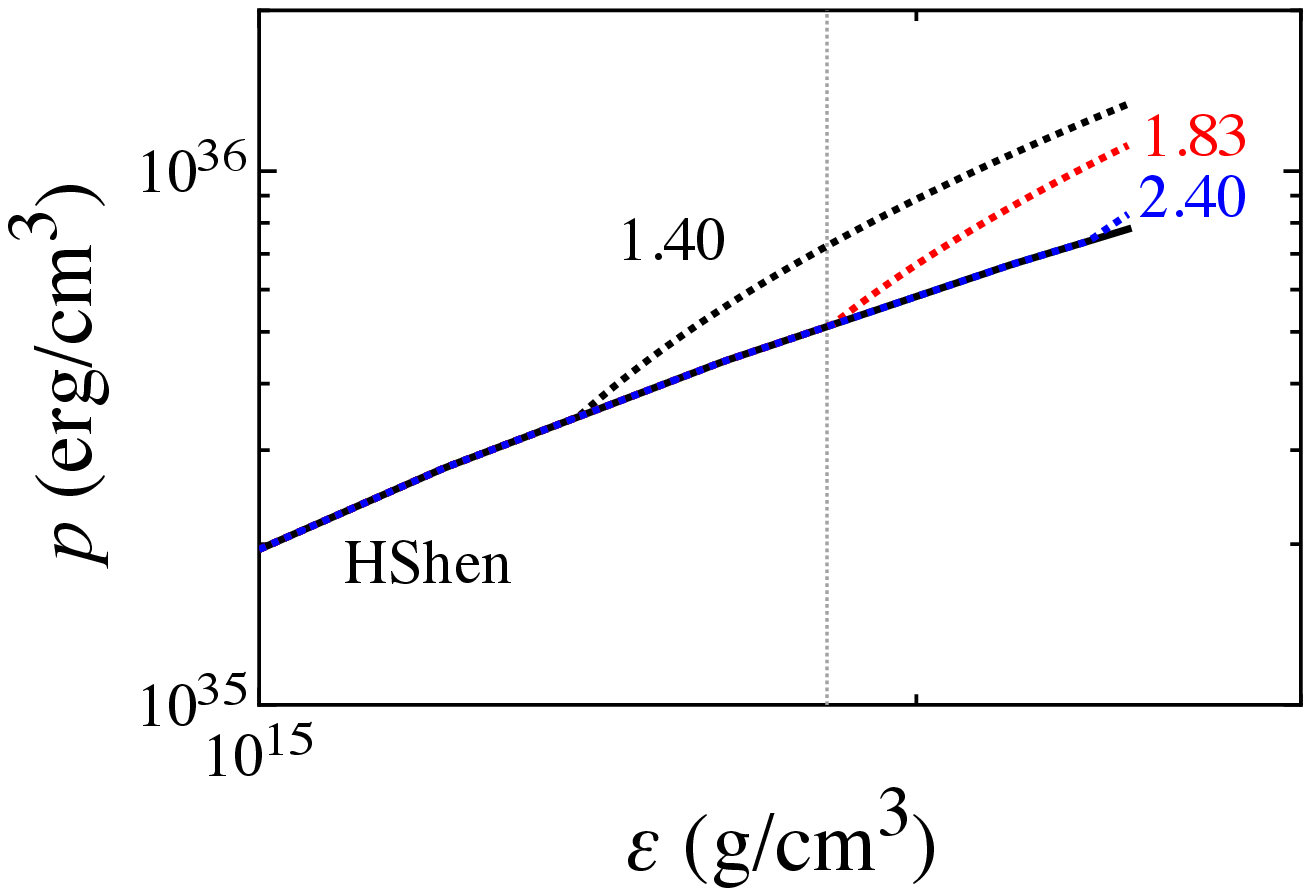} &
\includegraphics[scale=0.5]{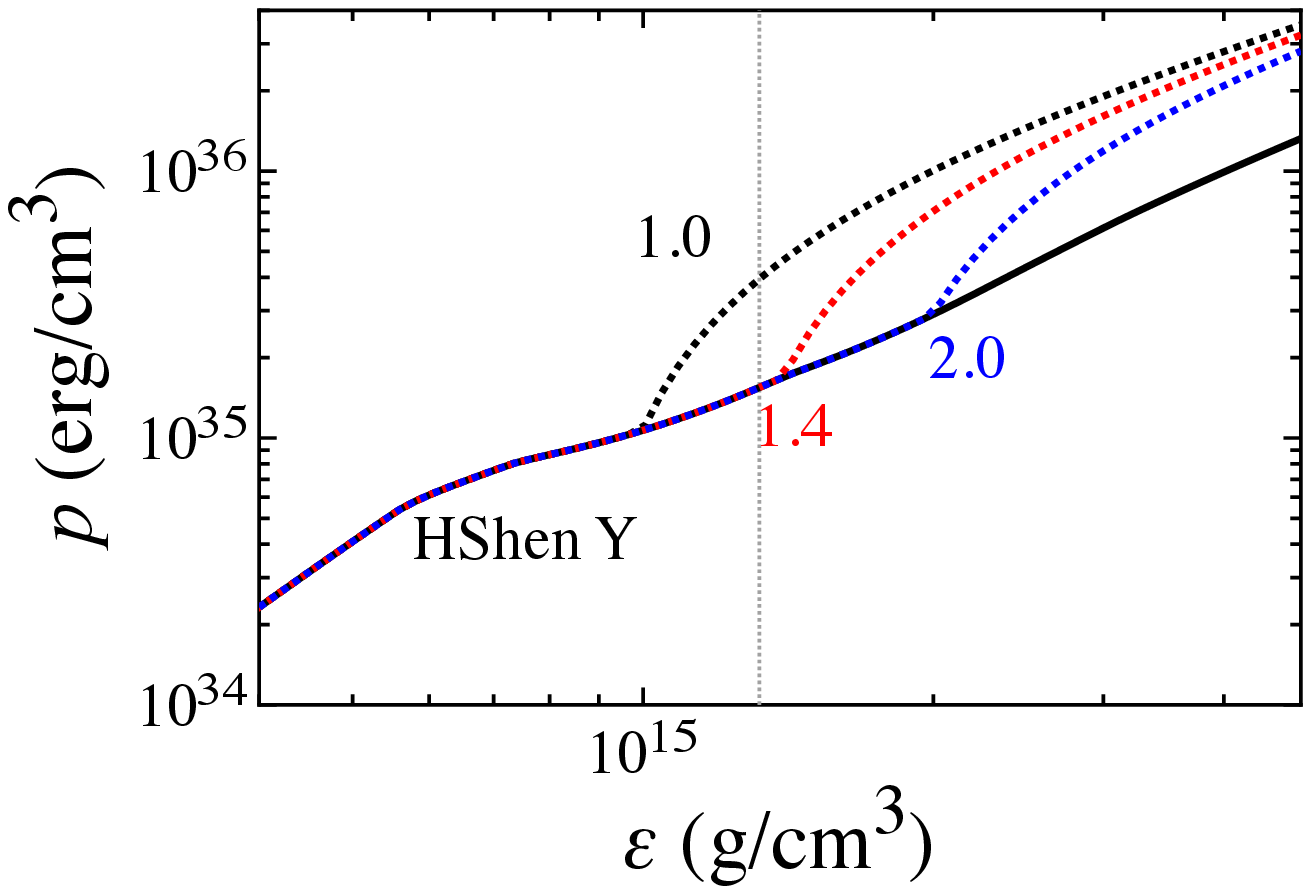}
\end{tabular}
\end{center}
\caption{
EOS for constructing hybrid stars. In this figure, as an example, we show the HShen EOS (left panel) and HShen Y EOS (right panel) as hadronic EOS, which are connected to the EOS for quark matter. The solid line corresponds to the HShen or HShen Y EOS, while the dotted lines correspond to the EOS for quark matter with different value of the transition density $\varepsilon_{\rm L}$ denoted in the unit of $10^{15}$ g/cm$^3$. For reference, the vertical lines denote the central density with which the mass of neutron star becomes maximum for each hadronic EOS, as shown in Table \ref{tab:NS-properties}. 
}
\label{fig:EOS-L}
\end{figure*}

Finally we directly connect the hadronic EOS with the Landau EOS at the transition density $\varepsilon_{\rm L}$, as shown in Fig. \ref{fig:EOS-L}, i.e., we set $\varepsilon_c = \varepsilon_{\rm L}$ in Eq. (\ref{hqeos}). In particular, in Fig. \ref{fig:EOS-L}, we show the HShen (left panel) and HShen Y EOSs (right panel) as an example, which are connected to the EOS describing  quark matter as in Eq. (\ref{eq:eos}). In this figure, the solid line corresponds to the HShen or HShen Y EOS, while the dotted lines are EOS for quark matter with different values of $\varepsilon_{\rm L}$ by changing the bag constant $\cal B$. The third family of compact objects is a sequence of the stable equilibrium configurations of compact objects, which are more compact than neutron stars, and the unstable equilibrium models of neutron stars should exist between the sequence of the third family of compact stars and that of the usual neutron stars. Thus, in definition, the transition density $\varepsilon_{\rm L}$ should be larger than the central density with which the neutron star mass constructed with the hadronic EOS becomes maximum. Note that the Landau EOS exhibits a strong stiffening after the phase transition.

\section{The third family of compact objects}
\label{sec:III}
\subsection{Hybrid stars as the third family of compact stars}

\begin{figure}
\begin{center}
\includegraphics[scale=0.5]{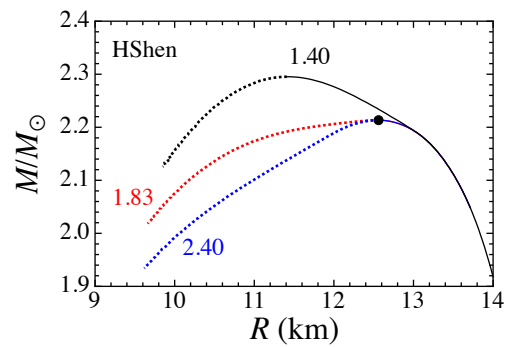}
\end{center}
\caption{
Mass-radius relations constructed with the HShen EOS connected to the Landau EOS at $\varepsilon_{\rm L}=1.40$, $1.83$, and $2.40\times 10^{15}$ g/cm$^3$. In the figure, the solid and dotted lines correspond to the stable and unstable equilibrium configurations, respectively. For reference, the maximum mass of neutron stars expected with the HShen EOS denotes by the filled circle in the figure.
}
\label{fig:MR-HShen}
\end{figure}

First, we consider hybrid stars with the HShen EOS, which is composed of only nucleons. In Fig. \ref{fig:MR-HShen}, we show the mass-radius relation for the cases of $\varepsilon_{\rm L}=1.40$, $1.83$, and $2.40\times 10^{15}$ g/cm$^3$, where the solid and dotted lines correspond to the stable and unstable configurations. For the case of $\varepsilon_{\rm L}=1.40\times 10^{15}$ g/cm$^3$, due to the stiffness of the Landau EOS, the maximum mass of hybrid star can become larger than that of neutron star constructed with the HShen EOS. However, since $\varepsilon_{\rm L}$ in this case is less than the central density with which the neutron star mass constructed with the HShen EOS becomes maximum, one can not see the third family of compact objects. For the other cases of $\varepsilon_{\rm L}=1.83$ and $2.40\times 10^{15}$ g/cm$^3$, the maximum masses are completely equivalent to that predicted from the HShen EOS, where one cannot observe the additional equilibrium sequence. That is, the third family of compact objects cannot be constructed by adopting the HShen EOS. We remark that the quark phase does not appear inside the stellar models for $\varepsilon_{\rm L}=1.83$ and $2.40\times 10^{15}$ g/cm$^3$, because the central density for the stable stellar model can not reach $\varepsilon_{\rm L}$. The reason why the third family of compact objects cannot appear for adopting the HShen EOS, may be the fact that the stellar mass constructed with the HShen EOS is so large that even the stiffness due to the Landau EOS cannot support.

Next, we consider the hybrid stars with the improved HShen EOS including the hyperons, i.e., the HShen $\Lambda$, HShen Y, and HShen Y$\pi$ EOSs. However, since the results with the HShen Y$\pi$ EOS are very similar to those with the HShen Y EOS, we remove the results with the HShen Y$\pi$ EOS from the following figures to avoid a complication. In Fig. \ref{fig:MR-HShenY}, we especially show the mass-radius relation for the cases of $\varepsilon_{\rm L}=1.0$, $1.4$, and $2.0\times 10^{15}$ g/cm$^3$ by adopting the HShen Y EOS as the hadronic EOS. In this figure, the solid and dotted lines correspond to the stable and unstable equilibrium configurations of the hybrid stars. From this figure, one can obviously observe the third family of compact objects for $\varepsilon_{\rm L}=1.4$ and $2.0\times 10^{15}$ g/cm$^3$. Unfortunately, the case of $\varepsilon_{\rm L}=2.0\times 10^{15}$ g/cm$^3$ has to be ruled out, because the expected maximum mass is less than $2M_\odot$, but the case of $\varepsilon_{\rm L}=1.4\times 10^{15}$ g/cm$^3$ can still survive the $2M_\odot$ problem. Additionally, from this figure, we find that the maximum mass of the third family of compact objects decreases as the value of $\varepsilon_{\rm L}$ increases. And, the third family of compact objects eventually becomes unstable when the value of $\varepsilon_{\rm L}$ is over a critical value.
By definition, the radius of the third family of compact object should be smaller than that of a usual neutron star, while the mass of the third family might be comparable to that of a usual neutron star. Thus, one possibility to be distinguish whether an object is the third family or usual neutron star is the observations of two objects with different radii but comparable masses, i.e., the so-called twin stars. Additionally, the careful observations of cooling history of compact objects tell us whether quark matter exists inside the star, because the cooling history depends on the existence of quark matter. This might be a hint to consider a possibility of the third family of compact objects.

\begin{figure}
\begin{center}
\includegraphics[scale=0.5]{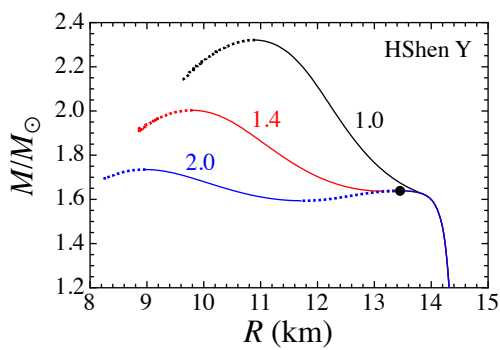}
\end{center}
\caption{
Mass-radius relations constructed with the HShen Y EOS connected to the Landau EOS at $\varepsilon_{\rm L}=1.0$, $1.4$, and $2.0\times 10^{15}$ g/cm$^3$. In the figure, the solid and dotted lines correspond to the stable and unstable equilibrium configurations, respectively. One can observe the third family of compact objects for the cases of $\varepsilon_{\rm L}=1.4$ and $2.0\times 10^{15}$ g/cm$^3$. For reference, the maximum mass of neutron stars expected with the HShen Y EOS denotes by the filled circle in the figure.
}
\label{fig:MR-HShenY}
\end{figure}

\begin{figure}
\begin{center}
\includegraphics[scale=0.5]{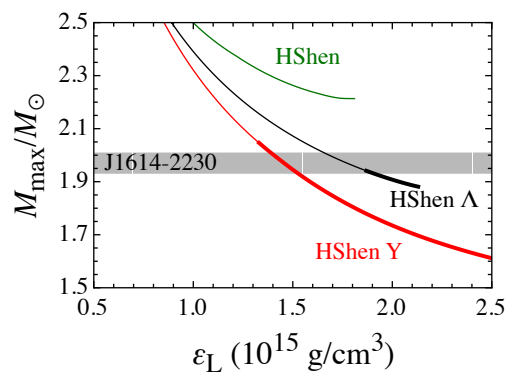}
\end{center}
\caption{
The maximum mass of star is shown as a function of the transition density $\varepsilon_{\rm L}$. The thick parts correspond to the third family of compact objects. For reference, the observational constraint on the mass of the millisecond pulsar J1614-2230 is also shown, which is $M=(1.97\pm 0.04)M_\odot$ \citep{D2010}.
}
\label{fig:M-eL}
\end{figure}

In Fig. \ref{fig:M-eL}, we show the maximum mass of hybrid stars as a function of $\varepsilon_{\rm L}$ for each hadronic EOS. In particular, the thick lines correspond to the third family of compact objects, where the value of $\varepsilon_{\rm L}$ is larger than the central density with which the mass of the neutron star constructed with the hadronic EOS becomes maximum. From this figure, one can see that hybrid stars with the improved HShen EOSs with hyperons can realize the third family of compact objects. However, considering the existence of the $2M_\odot$ neutron stars, the parameter space of $\varepsilon_{\rm L}$ and the hadronic EOS for realizing the third family of compact objects seem to be quite limited. In Fig. \ref{fig:M-eL}, for reference, we also add the observational evidence of massive neutron stars, whose mass is estimated to be $M=(1.97\pm 0.04)M_\odot$ \citep{D2010}. Due to the existence of such a massive neutron star, the third family of compact objects constructed with only the HShen Y EOS can survive, although the allowed parameter range of $\varepsilon_{\rm L}$ is not so large. Meanwhile, the third family constructed with the HShen $\Lambda$ is marginal. We remark that the third family with the HShen Y$\pi$ EOS can be produced, but the maximum mass of any compact object in the third family cannot reach $2M_\odot$.

Comparing the results obtained up to now with Fig. \ref{fig:HShen} and Table \ref{tab:NS-properties}, we may be able to extract the conditions 
to realize the third family of compact stars whose maximum mass becomes over $2M_\odot$: for the neutron star models constructed with the hadronic EOSs, the maximum mass and  the corresponding central density should be small, and the corresponding stellar radius should be large. Thus, we remark that there might be a possibility that the soft EOSs with hyperons and/or the other components which have been ruled out after the discovery of the $2M_\odot$ neutron star, could revive owing to the shift to the third family of compact objects.

\begin{figure}
\begin{center}
\includegraphics[scale=0.5]{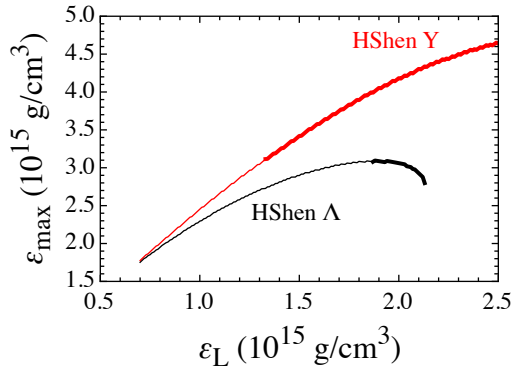}
\end{center}
\caption{
The central density with which the mass of the star becomes maximum is shown as a function of the junction density $\varepsilon_{\rm L}$. The thick parts correspond to the third family of compact objects. 
}
\label{fig:emax-eL}
\end{figure}

\begin{figure}
\begin{center}
\includegraphics[scale=0.5]{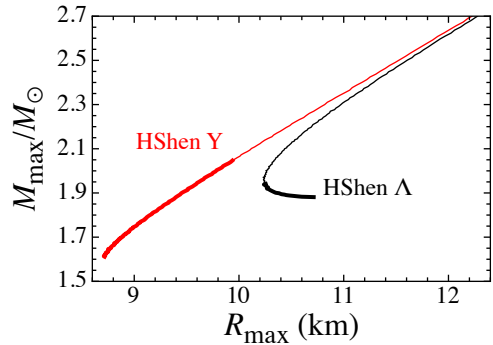}
\end{center}
\caption{
The maximum mass of star is shown as a function of the stellar radius for the hybrid star with the maximum mass. The thick parts correspond to the third family of compact objects. 
}
\label{fig:MmaxRmax}
\end{figure}

\begin{figure}
\begin{center}
\includegraphics[scale=0.5]{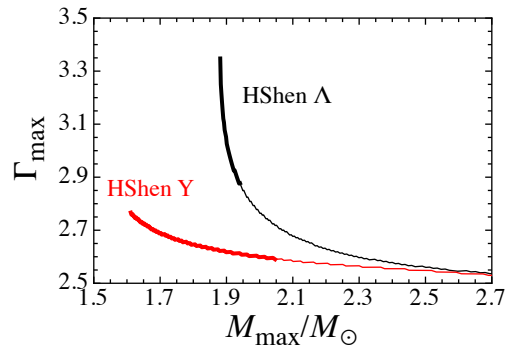}
\end{center}
\caption{
Adiabatic index at the central density for the hybrid star with maximum mass is shown as a function of the maximum mass of star. The thick parts correspond to the third family of compact objects.
}
\label{fig:Gamma}
\end{figure}

We discuss some properties of the third family of compact objects with the additional figures. In Fig. \ref{fig:emax-eL}, we show the central density with which the hybrid star becomes maximum mass, $\varepsilon_{\max}$, as a function of $\varepsilon_{\rm L}$ for the HShen $\Lambda$ and HShen Y EOSs, where the thick lines correspond to the third family of compact objects. From this figure, we find that the value of $\varepsilon_{\max}$ does not monotonically increase with $\varepsilon_{\rm L}$. In Fig. \ref{fig:MmaxRmax}, we show the maximum mass of hybrid stars as a function of the stellar radius when the mass of hybrid star becomes maximum, $R_{\rm max}$, as changing the value of $\varepsilon_{\rm L}$ for the HShen $\Lambda$ and HShen Y EOSs. One can see that the stellar radius $R_{\rm max}$ and the maximum mass of hybrid star generally decrease as the value of $\varepsilon_{\rm L}$ increases up to a critical value. In Fig. \ref{fig:Gamma}, we show the adiabatic index at the stellar center for the maximum mass hybrid star, $\Gamma_{\rm max}$, as a function of the maximum mass of hybrid star by changing the value of $\varepsilon_{\rm L}$, where the adiabatic index $\Gamma$ is defined as
\begin{equation}
  \Gamma = \frac{p+\varepsilon}{p}\left(\frac{dp}{d\varepsilon}\right).  \label{eq:gamma}
\end{equation}
As the value of $\varepsilon_{\rm L}$ increases and reaches the critical value with which the third family of compact objects becomes unstable equilibrium, the value of $\Gamma_{\rm max}$ also becomes maximum, although such stellar models are ruled out from the $2M_\odot$ neutron star observation.

\subsection{Criteria for the third family of compact stars}

\begin{figure*}
\begin{center}
\begin{tabular}{cc}
\includegraphics[scale=0.5]{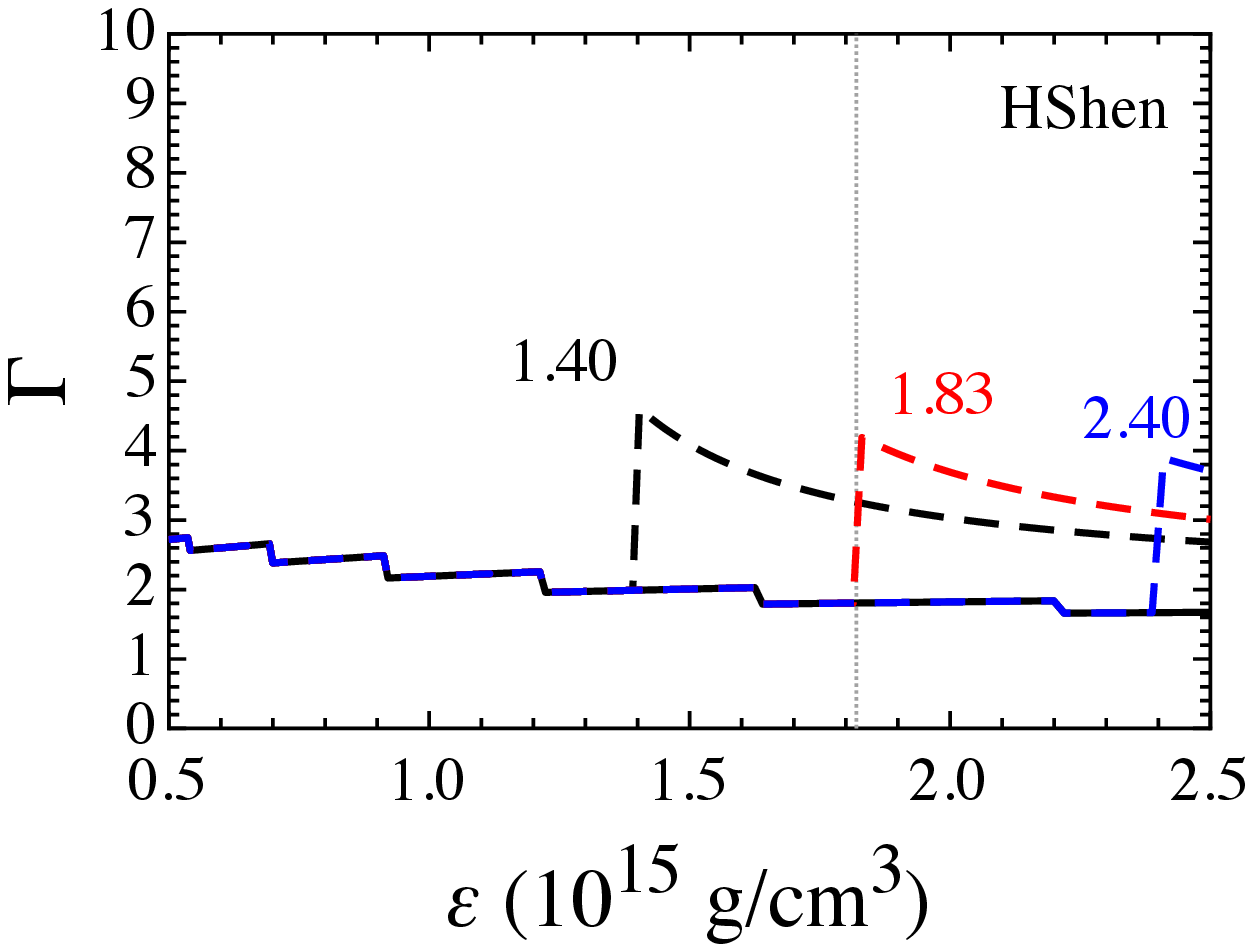} &
\includegraphics[scale=0.5]{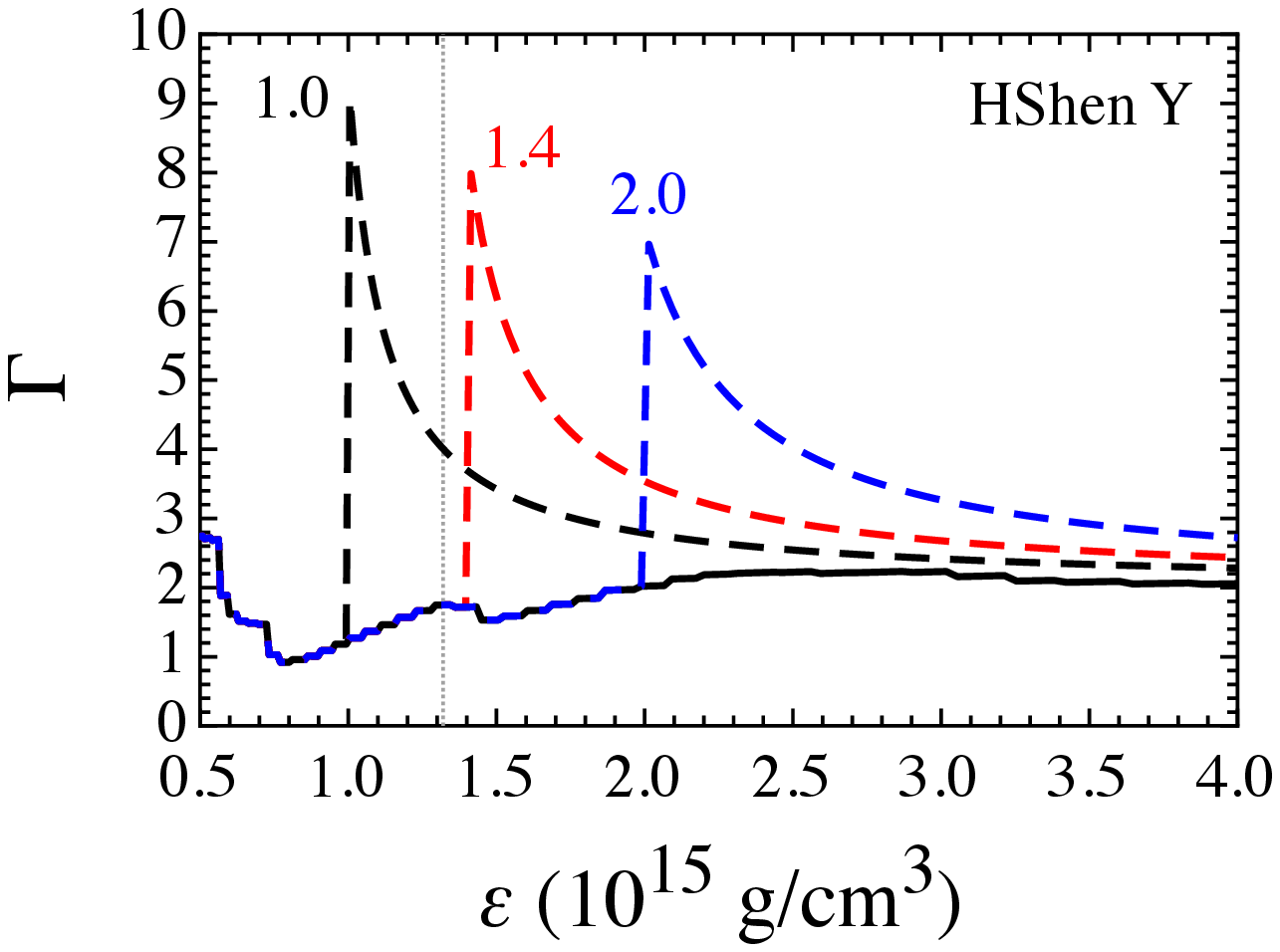}
\end{tabular}
\end{center}
\caption{
With the HShen and HShen Y EOSs, the adiabatic indexes are shown as a function of the energy density, where the solid and broken lines correspond to the results for the stellar models without and with the phase transition into quark matter. The labels in the figure denote the transition density $\varepsilon_{\rm L}$ in the unit of $10^{15}$ g/cm$^3$. From this figure, one can obviously observe that the jump of adiabatic index with HShen EOS at the transition density is much smaller than the case with HShen Y EOS. For reference, the vertical lines denote the central density with which the mass of neutron star constructed with each hadronic EOS becomes maximum.
}
\label{fig:gamma-e}
\end{figure*}

\begin{figure}
\begin{center}
\includegraphics[scale=0.5]{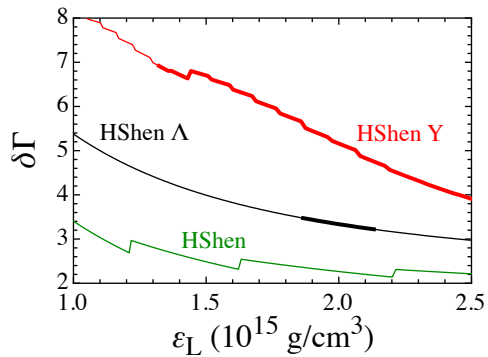} 
\end{center}
\caption{
The difference of the adiabatic indexes between the stellar models without and with the phase transition into quark matter at the transition density $\varepsilon_{\rm L}$, are shown as a function of $\varepsilon_{\rm L}$. The thick part on each line corresponds to the third family of compact objects. 
}
\label{fig:dgamma}
\end{figure}

Now, we would like to give a conjecture about the conditions when the third family of compact object exists. As mentioned before, at least the central density of the compact object should be larger than that for the neutron star with the maximum mass. In addition, in order to recover the stability, EOS for the inner region where the phase transition arises from the hadronic into quark matter, must be significantly stiff. To check this point, we focus on the adiabatic index calculated with Eq. (\ref{eq:gamma}), i.e.,
\begin{equation}
\Gamma=\frac{2(\varepsilon-{\cal B})}{\varepsilon-2{\cal B}},
\end{equation}
for the Landau EOS. Then, we can observe that $\Gamma$ becomes very large at the transition density, since $\varepsilon_{\rm L}$ is relatively close to $2\cal B$. Meanwhile, $\Gamma$ approaches the asymptotic value of 2 in the high-density region. As an example, in Fig. \ref{fig:gamma-e} we show the adiabatic index as a function of the energy density especially with the HShen and HShen Y EOSs, where the solid and broken lines correspond to the adiabatic indexes for compact objects without and with the phase transition into quark matter. In the figure, the numbers denote the transition density $\varepsilon_{\rm L}$ in the unit of $10^{15}$ g/cm$^3$. From this figure, one can observe that the adiabatic index for compact objects with the phase transition increases abruptly at $\varepsilon=\varepsilon_{\rm L}$, and that the difference between the adiabatic indexes without and with the phase transition decreases as $\varepsilon_{\rm L}$ increases. On the other hand, as shown before, the compact object cannot recover the stability, when the value of $\varepsilon_{\rm L}$ exceeds a critical value. From these results, we consider that the difference between the adiabatic indexes without and with the phase transition should be larger than a critical value, if the third family of compact objects exist. So, in Fig. \ref{fig:dgamma}, we plot the difference $\delta\Gamma$ as a function of $\varepsilon_{\rm L}$ for the HShen, HShen $\Lambda$, and HShen Y EOSs. Here, $\delta\Gamma$ is defined as
\begin{equation}
  \delta \Gamma = \Gamma_{\rm HS} - \Gamma_{\rm NS},
\end{equation}
where $\Gamma_{\rm HS}$ and $\Gamma_{\rm NS}$ denote the adiabatic indexes at $\varepsilon=\varepsilon_{\rm L}$ for the compact stars with and without quark matter. In this figure, the thick part on each line corresponds to the third family of compact object. From this figure, we find that the large jump of $\Gamma$ at the phase transition is necessary so that the third family exists. In fact, $\delta \Gamma$ for the HShen EOS is less than $\sim 3$ for the central density larger than that for the neutron star with the maximum mass. This may be a reason why the third family can not exist for the HShen EOS. Namely, the condition for existence of the third family might be that $\delta\Gamma$ should be larger than a critical value for the stellar models constructed with a central density larger than that for the neutron star with the maximum mass, where the exact value of such a critical value of $\delta\Gamma$ might depend on the hadronic EOS (see Appendix \ref{sec:appendix_1}).

It should be interesting to compare our results with the ordinary bag model EOS expressed by Eq.~(\ref{bagEOS}). The transition density becomes very high in this case, and there is no transition for the large bag constant, for example ${\cal B}=340$ MeV/fm$^{3}$ with $\varepsilon_{\rm L}=1.4\times 10^{15}$ g/cm$^3$ for the HShen Y EOS. The corresponding adiabatic index can be written as 
\begin{equation}
   \Gamma=\frac{4}{3}\frac{\varepsilon-{\cal B}}{\varepsilon-4{\cal B}}. 
\end{equation}
Thus, we can find both effects the transition density and $\delta\Gamma$ disfavor the existence of the third family in the ordinary bag model, besides the stability condition [Eq. (\ref{cond1})].

{\subsection{Phase transition and the third family}

Finally, we would like to make some comments about the relation between the phase transition and the possible existence of the third family. In  \cite{Kamp1981a,Kamp1981b,Glen2000b, Schaf2002}, they have also discussed the possibility of the appearance of the third family following the phase transitions. One of the points discussed there is that the softening of EOS first occurs at the transition point and then the stiffening follows at high-density. That is, the adiabatic index abruptly decreases and cannot satisfies the gravitational stability condition [Eq.~(\ref{cond1})] just after the phase transition, which leads stars to be gravitationally unstable. In this case, the phase transition must occurs at relatively low density and accordingly the gravitational instability occurs before reaching the maximum mass constructed with the normal EOS, which is the EOS without the phase transition. This is a result owing to the appearance of the mixed phase \citep{Glen2000b, Schaf2002}. After that, EOS becomes stiff again at higher densities due to the additional effects in the new phase. The transition from the mixed phase to the pure quark phase may give rise to an increase of the adiabatic index, which is similar to our model for the hadron-quark phase transition to generate the third family \citep{Glen2000b}. However, the maximum mass of the third family discussed in \cite{Glen2000b} and in \cite{Schaf2002} cannot be large and probably less than the corresponding maximum mass constructed with normal EOS; as already stated the adiabatic index of the pure quark phase cannot become too large in the case of the hadron-quark phase transition. In other words, if we discard the appearance of the mixed phase in the models of \cite{Glen2000b} and of \cite{Schaf2002}, we have no third family and we only have the mass-radius relation with maximum mass less than that with normal EOS. These features are quite different from our results. We have seen the large increase of the adiabatic index at the transition point due to the effect of magnetic field, which generates the third family of compact stars. We must point out at the same time that the adiabatic index approaches to two at high-density region, which supports the large maximum mass of hybrid stars.

We have not taken into account the details of the phase transition such as the mixed phase. One may consider a discontinuity of the energy density $\Delta\varepsilon$ at the transition point in Eq.~(\ref{hqeos}),
\begin{equation}
  p(\varepsilon)=\left\{
  \begin{array}{@{\,}cl@{\,}}
     p_H(\varepsilon) & \varepsilon < \varepsilon_c \\
     p_Q(\varepsilon) & \varepsilon \ge \varepsilon_c + \Delta\varepsilon
  \end{array}\right. ,
\end{equation}
with $\varepsilon_c$ determined by $p_H(\varepsilon_c)=p_Q(\varepsilon_c+\Delta\varepsilon)$, which may resemble the Maxwell construction in the case of the first order phase transition \citep{AMP2013}. The EOS is then softened near the transition density by the introduction of $\Delta\varepsilon$, which gives rise to a gravitational instability around the onset of the third family in Fig. \ref{fig:MR-HShenY}. However, such instability is soon recovered by the large $\delta\Gamma$. The appearance of the mixed phase also soften the EOS near the transition density to diminish the stable branch of the third family, if the mixed state is taken into account. Anyway, the bulk properties such as the maximum mass are little changed even in that case \citep{ST2015}.

\section{Conclusion and Discussion}
\label{sec:IV}

In this paper, we have considered the possibility for existence of the third family of compact objects, which are stable equilibrium configurations more compact than neutron stars. It is well known that white dwarfs are supported by the electron degeneracy pressure, and neutron stars by the neutron degeneracy pressure together with the strong nuclear force in the short range. However, it is not sure whether the stable objects more compact than neutron stars can exist or not. To construct such compact objects, i.e., the third family of compact objects, one has to take into account the additional physics to support the gravity generated by such objects. As additional physics, we have considered the magnetic field inside hybrid stars, whose strength and geometry are not fixed observationally. If the magnetic fields inside the star would be so strong that quarks settle only in the lowest Landau level, the equation of state for quark matter becomes quite stiff \citep{ST2015}. 
To realize such a stiff EOS, the magnetic field strength should become $\sim {\cal O}(10^{19} {\rm G}$), where the quark core of hybrid star becomes $\sim (70-90) \%$ of stellar radius\footnote{We consider the strong magnetic field inside the quark core for satisfying the condition that quarks settle on only the lowest Landau level. If such a strong magnetic field approaches the stellar surface although it may not be realistic, one could expect to observe a kind of extremely active electromagnetic phenomena from the stellar surface.}.
With such an extreme case, we have examined the possibility whether the third family of compact objects can exist.

Then, we have shown that the third family of compact objects can exist, even though the relevant parameter space is not so large and EOS for hadronic matter might be restricted, considering the observational constraint on the stellar mass such as $2M_\odot$. If such stars exist, they are hybrid stars composed of primarily quarks. By way of the numerical examinations, we have found that, for constructing the third family of compact objects, the hadronic EOSs could be favored, with which the maximum mass of neutron star should be small with a small central density and a large radius. That is, considering the possibility for the existence of third family of compact objects, the hadronic EOSs including hyperons may not be ruled out, even if the maximum mass of neutron star constructed with such soft hadronic EOSs is predicted to be less than $2M_\odot$. We additionally remark that, if the third family exists, the compact objects whose radii are smaller than neutron stars with the same mass can exist even in the narrow mass range, i.e., the existence of twin stars. Furthermore, if compact objects in the third family is formed in supernovae with larger masses than the maximum mass of neutron stars, one might expect to observe the second bounce when the central density reaches the transition density where quark matter appears, which is similar to the usual bounce in supernovae arising when the central density reaches the nuclear saturation density. As another possibility, one might distinguish compact objects in the third family from usual neutron stars by observing the thermal evolution, because the cooling history could strongly depend on the interior compositions.

In this paper, we have used the simple EOS for quark matter, based on the MIT bag model. For more realistic description, other EOS, e.g., based on the perturbative QCD or the NJL model might be desirable; especially the latter is appropriate to include chiral dynamics, which is a basic concept of underlying QCD. Actually the inhomogeneous chiral phase has been actively studied in the presence of the magnetic field \citep{Bub2015,TNK2015,NKT2015,YNT2015}. It should be interesting to take into account such aspect in the next step. We have adopted the simple procedure to connect the quark EOS with the hadronic EOS. Thus, we should make an additional analysis, where we might have to consider the physical connection between the EOSs for quark and hadronic matters. Moreover, it might be more realistic to take into account the effects of magnetic field even in hadronic matter. Such additional analyses would be done somewhere in future.

We have seen that EOS is strongly stiffened after the deconfinement transition due to the strong magnetic field, and the adiabatic index becomes larger than the critical value of the gravitational stability, $4(1+KR_s/R)/3$. However, only the large adiabatic index in EOS is not sufficient for the possible existence of the third family. The transition density must be not so large, and at the same time the adiabatic index following the phase transition must increase enough to large: a combination of both conditions can produce the third family. These observations may partially support the conjecture suggested in \cite{G1968}. Additionally, we have neglected the magnetic pressure for constructing the stellar models. At least, the poloidal magnetic fields can increase the maximum mass of the stars, which may be an advantage for realizing the third family of compact objects. Such effects would be also considered somewhere in future. Furthermore, it should be noted that the magnetic pressure might depend on the geometry of magnetic fields\footnote{\cite{Franzon2016} considered the magnetic geometry in neutron stars constructed with EOS including effects of magnetic field consistently.}. \cite{HHRS2010} suggested the possibility that the transverse pressure tends to vanish due to the freezing of matter in the lowest Landau level if the magnetic field is so strong. Meanwhile, \cite{Ferrer2010} suggested the possibility of a pressure anisotropy due to the strong magnetic field, while Potekhin and Yakovlev pointed out that the pressure in the magnetized objects is independent of the magnetic direction \citep{PY2012}. If such an anisotropic pressure really appears under the strong magnetic field, one might have to consider such effects on the neutron star structure. Anyhow, the anisotropic pressure should not be so serious yet for the magnetic field of ${\cal O}(10^{19}$ G) \citep{HHRS2010}.

We are grateful to C. Ishizuka for preparing some EOS data and to A. Ohnishi for discussing the EOSs adopted in this paper. This work was supported in part by Grants-in-Aid for Scientific Research on Innovative Areas through 
No.\ 24105008 and No.\ 15H00843 provided by MEXT, and by Grant-in-Aid for Young Scientists (B) through No.\ 26800133 provided by JSPS.

\appendix
\section{Dependence of $\delta \Gamma$ on the hadronic EOS}   
\label{sec:appendix_1}

We adopt only a few EOSs in this paper to consider the third family of compact objects, because there are a few available EOSs with hyperon predicting that the stellar radius of the neutron star model with maximum mass is relatively larger. Even so, in this appendix we propose a speculation which hadronic EOS can produce the third family. For this purpose, we have to know the critical value of $\delta\Gamma$ for each EOS, i.e., the minimum value of $\delta \Gamma$, $\delta\Gamma_{\rm min}$, with which the third family exists. As shown in Fig. \ref{fig:dgamma}, $\delta\Gamma_{\rm min}$ for HShen $\Lambda$  EOS corresponds to the right edge of the thick line. To determine $\delta \Gamma_{\rm min}$ for HShen Y and HShen Y$\pi$ EOSs, we consider the region even for ${\cal B}\gsim 500$ MeV/fm$^3$. Then, we are successful to find the strong correlation between $\delta \Gamma_{\rm min}$ and the radius of the neutron star with maximum mass. In Fig. \ref{fig:dg-R}, we show the values of $\delta \Gamma_{\rm min}$ as a function of the radius of the neutron star with the maximum mass for three EOSs, i.e., the HShen $\Lambda$, HSHen Y, and HShen Y$\pi$ EOSs. In this figure, we also plot the fitting formula with the dotted line, which is given by
\begin{equation}
  \delta\Gamma_{\rm min} = 5.583 - 1.895\left(\frac{R}{10 \,{\rm km}}\right). \label{eq:critical}
\end{equation}
The region above the dotted line corresponds to the region where the third family exists. That is, if one prepares a hadronic EOS, one can determine the stellar radius of the neutron star with maximum mass and the $\delta \Gamma$ for the stellar model with the transition density which is exactly equal to the central density for the neutron star with the maximum mass. If such a set of the radius and $\delta \Gamma$ is plotted in the region below the line given by Eq. (\ref{eq:critical}), the third family might not exist with the adopted hadronic EOS, because $\delta \Gamma$ would basically decrease as the transition density increases (see Fig. \ref{fig:dgamma}) and $\delta\Gamma$ determined here should be almost maximum value of $\delta \Gamma$. For example, with HShen EOS, such radius and $\delta \Gamma$ become $R=12.56$ km and $\delta \Gamma=2.39$, which is in the region below the line given by Eq. (\ref{eq:critical}). Anyway, to derive the critical line, we should check the correlation between $\delta \Gamma_{\rm min}$ and some properties in hadronic EOS, using more samples.

\begin{figure}
\begin{center}
\includegraphics[scale=0.5]{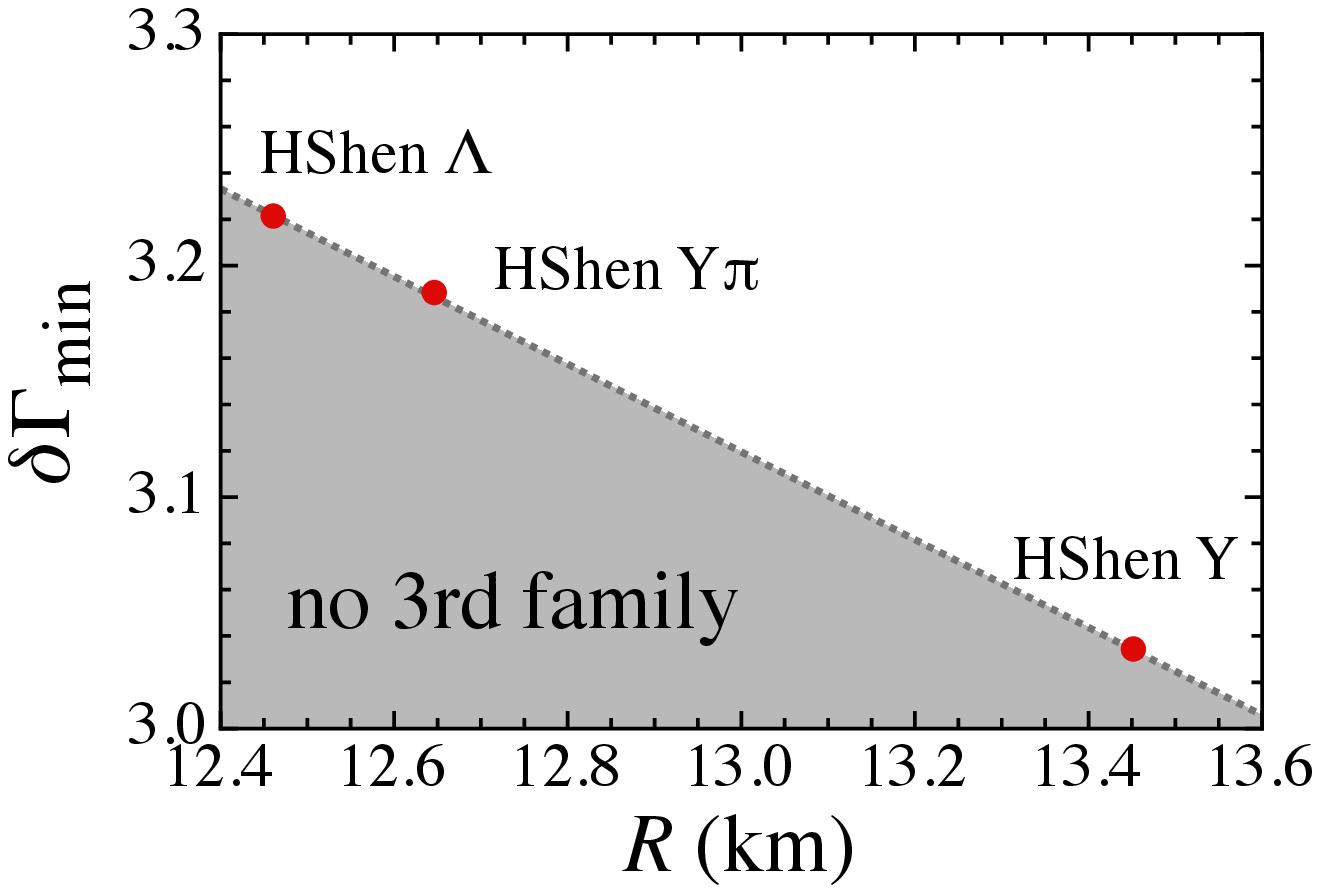}
\end{center}
\caption{
The minimum value of $\delta\Gamma$ are plotted as a function of the radius of the neutron star with the maximum mass. 
}
\label{fig:dg-R}
\end{figure}


\end{document}